\documentclass[12pt]{article}
\usepackage{graphicx}
\usepackage{amssymb}
\usepackage{amsmath}
\usepackage{epsfig}

\newcommand{\singlespace}{\renewcommand{\baselinestretch}{1}\large\normalsize}
\newcommand{\doublespace}{\renewcommand{\baselinestretch}{1.6}\large\normalsize}

\newcommand{\beq}{\begin{equation}}
\newcommand{\eeq}{\end{equation}}
\newcommand{\bea}{\begin{eqnarray}}
\newcommand{\eea}{\end{eqnarray}}
\newcommand{\ave}[1]{\langle {#1} \rangle}

\newcommand{\dslash}{\partial\!\!\!/}

\newcommand{\eq}[1]{Eq.~(\ref{#1})}

\def\roughly#1{\mathrel{\raise.3ex\hbox{$#1$\kern-.75em%
\lower1ex\hbox{$\sim$}}}}
\def\lsim{\roughly<}

\def\={\;=\;}
\def\+{\;+\;}

\def\bear{\begin{array}}
\def\ear{\end{array}}

\def\newline{\hfil\break}

\begin{document}

\begin{center}
\doublespace
\begin{large}
{\bf } Flavor-Mixing Effects on the QCD Phase Diagram at non-vanishing        
       Isospin Chemical Potential: One or Two Phase Transitions?
\end{large}
\vskip 0.4in

M. Frank$^{a,b}$, M. Buballa$^{a,c}$, and M. Oertel$^{d}$

{\small{\it 
            $^a$ Institut f\"ur Kernphysik, TU Darmstadt,
                 Schlossgartenstr. 9, 64289 Darmstadt, Germany\\
            $^b$ Fachbereich Mathematik, TU Darmstadt,
                 Schlossgartenstr. 7, 64289 Darmstadt, Germany\\
            $^c$ Gesellschaft f\"ur Schwerionenforschung (GSI),
                 Planckstr. 1, 64291 Darmstadt, Germany\\
            $^d$ IPN-Lyon, 43 Bd du 11 Novembre 1918,
                 69622 Villeurbanne C\'edex, France}}\\
\end{center}

\vspace{8mm}

\begin{abstract}
We investigate effects of a fixed nonzero isospin chemical potential on the
$\mu_B$-$T$ phase diagram of strongly interacting matter using a
Nambu--Jona-Lasinio-type four fermion interaction. We focus on the
influence of a flavor-mixing interaction induced by instantons. We
find that already for rather moderate values of the coupling strength in
the flavor-mixing channel the recent findings of two seperate phase
transitions do not persist.
\end{abstract}

\singlespace
The structure of the phase diagram of QCD at nonzero temperature and
baryon chemical potential has been intensively studied throughout the
last decade, as well on theoretical as on experimental
side. In the baryon chemical potential-temperature ($\mu_B$-$T$) plane
roughly speaking three domains can be distuingished: the hadronic
phase at low temperature and density, a quark matter phase at low
temperature and high baryon density, which is most probably a color
superconductor~\cite{RaWi00,Alford01}, and the quark gluon plasma at high
temperature. For two flavors at zero chemical potential the hadronic
phase is expected to be seperated by a crossover from the
quark-gluon-plasma~\cite{K01}. The fact that at zero temperature a 
first-order transition to quark matter is expected~\cite{HJSSV98} implies the
existence of a second-order endpoint somewhere in the
$\mu_B$-$T$-plane. This second-order endpoint is one of the features
of the QCD phase diagram which could be detected in heavy-ion
collisions~\cite{SRS99}. 

Up to now most theoretical investigations of the
phase diagram and  are restricted to zero
isospin chemical potential $\mu_I$ (see, e.g., Refs.\cite{RaWi00,Alford01,
HJSSV98,SRS99}). The influence of a non-zero $\mu_I$ has been
studied in more detail in the context of quark matter at high baryon density
and low temperature, e.g.,~\cite{Bedaque02,ABR01,KYT02}, 
as well as for vanishing $\mu_B$~\cite{SoSt01,KoSi02}. 

Recently the $\mu_B$-$T$ phase diagram for a fixed $\mu_I
\neq 0 $ has been studied using a random matrix
model~\cite{KTV03} and a Nambu--Jona-Lasinio (NJL)
model~\cite{TK03}. The authors of Refs.~\cite{KTV03,TK03} find
striking effects: There are two first-order phase transitions at
low temperature and high baryon chemical potential and thus two 
second-order endpoints. This could be important, e.g., for heavy-ion
collisions where $\mu_I$ is supposed to be nonzero. In this letter we
will argue that the existence of two separate phase transitions becomes
unlikely, once flavor-mixing effects due to instantons are taken into
account. 

Our starting point is the following Lagrangian for two flavors: 
\beq
{\cal L} = {\cal L}_0 + {\cal L}_1 + {\cal L}_2~, 
\eeq 
with a free part
\beq 
{\cal L}_0 = {\bar q} ( i \dslash - m ) q~, 
\eeq 
and two
different interaction parts (see, e.g., \cite{AsYa89,Klevansky}),
\beq 
{\cal L}_1 = G_1 \Big\{ ({\bar q}
q)^2 + ({\bar q}\,\vec\tau q)^2 + ({\bar q}\,i\gamma_5 q)^2 + ({\bar
q}\,i\gamma_5\vec\tau q)^2 \Big\} 
\eeq 
and 
\beq 
{\cal L}_2 = G_2
\Big\{ ({\bar q} q)^2 - ({\bar q}\,\vec\tau q)^2 - ({\bar
q}\,i\gamma_5 q)^2 + ({\bar q}\,i\gamma_5\vec\tau q)^2 \Big\}~, 
\eeq
where  $ q = (u,d)^T$.
$G_1$ and $G_2$ are coupling contants of dimension $energy^{-2}$ and
$m = diag(m_u,m_d)$ contains the current quark masses. For simplicity we
assume $m_u = m_d$. The interaction is invariant under
$SU_L(2)\times SU_R(2)\times U_V(1)$ transformations. ${\cal L}_1$
exhibits an additional $U_A(1)$ symmetry, whereas this is not true for
${\cal L}_2$. ${\cal L}_2$ has the structure of a 't-Hooft determinant
in flavor space~\cite{tHooft}. This interaction can be interpreted as 
induced by instantons and reflects the $U_A(1)$-anomaly of QCD. 
The reason for this particular choice of the two interaction parts will 
become clear below.

We now want to study the properties of the system at temperature $T$
and the up and down quark chemical potentials
\beq
\mu_u = \mu + \delta\mu~,\quad \mu_d = \mu - \delta\mu~,
\label{mudef}
\eeq
which are in general different. Here $\mu = \mu_B/3$ is the quark number
chemical potential and $\delta\mu = \mu_I/2$ is proportional to the 
isospin chemical potential~\footnote{This follows from the definitions
of the baryon density, $n_B = \frac{1}{3}(n_u + n_d)$, and the 
isospin density $n_I = \frac{1}{2}(n_u-n_d)$ together with the 
thermodynamic relation $n_i = -\partial\Omega/\partial\mu_i$ for 
$i = u, d, B, I$. Note that often different definitions are used.
For instance, $\mu_B$ and $\mu_I$ of Refs.~\cite{KYT02,TK03} 
correspond to our $\mu$ and $\delta\mu$.}.   
To obtain the mean field thermodynamic potential $\Omega(T,\mu,\delta\mu)$,
we linearize the Lagrangian in the presence of
the following quark condensates:
\beq 
\phi_u = \ave{\bar u u}~,\qquad
\phi_d = \ave{\bar d d}~, 
\eeq 
which {\it a priori} can be different.  We do not consider $\rho =
\frac{1}{2} (\ave{\bar u \gamma_5 d} - \ave{\bar d \gamma_5 u})$, i.e.,
we exclude the possibility of pion condensation. This restricts our 
model to values of $|\delta\mu| < m_\pi/2 \simeq 70$~MeV. 
For the moment we also do not consider color superconducting phases.
This will, however, not change our results qualitatively.

For our further proceeding it is convenient to introduce constituent
quark masses:
\beq
     M_i = m_i - 4\,G_1\,\phi_i - 4\,G_2\,\phi_j~,\qquad i\neq j \in \{u,d\}~.
\label{masses}
\eeq
Using this definition and \eq{mudef}
for the chemical potentials of up and down quarks,
respectively, we obtain for the mean field thermodynamic potential:
\beq
     \Omega(T,\mu_u,\mu_d) = \sum_{f=u,d} \Omega_0(T,\mu_f;M_f)
     + 2\,G_1\,(\phi_u^2 + \phi_d^2) + 4\,G_2\phi_u\phi_d~,
\label{omega}
\eeq
where $\Omega_0(T,\mu_f;M_f)$ corresponds to the contribution of a gas
of quasiparticles of flavor $f$:
\begin{alignat}{2}
    \Omega_0(T,\mu_f;M_f) = -\frac{3}{\pi^2}\int dp\,p^2\,\Big\{
    E_f &+ T\,\ln{\Big(1 + \exp{(-\frac{1}{T}(E_f-\mu_f))}\Big)} \nonumber \\
        &+ T\,\ln{\Big(1 + \exp{(-\frac{1}{T}(E_f+\mu_f))}\Big)}\Big\}
\end{alignat}
with $E_f = \sqrt{M_f^2 + p^2}$. In order to determine the physical
solutions, we have to look for the stationary points of the
thermodynamic potential with respect to the two condensates $\phi_u$ and
$\phi_d$.  This leads to the following gap equations which have to be
solved selfonsistently:
\beq 
\phi_f = - \frac{3}{\pi^2} \int dp\,p^2\,
\frac{M_f}{E_f}\Big\{ 1 - n(E_f) - \bar{n}(E_f) \Big\}~,
\eeq
where $n(E_f) = 1/(\exp((E_f-\mu_f)/T)+1)$ and 
$\bar{n} = 1/(\exp((E_f+\mu_f)/T)+1)$ are Fermi occupation numbers.
Here we have to keep
in mind that via Eq.~(\ref{masses}), the constituent mass $M_i$ for one
flavor depends in general on both condensates and therefore the two
flavors are coupled. At this point our separation of the interaction
part becomes clear: If we for the moment neglect ${\cal L}_2$, i.e.,
$G_2 = 0$, the two flavors decouple. That means $M_i$ only
depends on the condensate of the same flavor $\phi_i$ and the mixed
contribution to $\Omega$ (last term of Eq.~(\ref{omega})) vanishes. In
this limit we recover the expression for the thermodynamic potential
of Ref.~\cite{TK03} for $\rho = 0$. In the opposite limit, i.e., $G_1
= 0$, we have ``maximal'' mixing: The constituent mass of flavor $i$
only depends on the condensate $\phi_j$ with $i \neq j$. 
For $G_1 = G_2$ we recover the original Lagrangian
proposed by Nambu and Jona-Lasinio~\cite{NJL}, implying $M_u = M_d$. 

To study the effects of flavor mixing, let us now write
\beq G_1 = (1-\alpha)\,G_0~,\qquad
G_2 = \alpha\,G_0~, 
\eeq
and calculate the phase diagram for fixed $G_0$ but different values of 
$\alpha$. The amount of flavor mixing is thereby controlled by the
particular value of $\alpha$ while the values of the vacuum
constituent quark masses $M_{vac}$ are kept constant.

For our numerical studies we use the following set of parameters: $m_u
= m_d = 6$~MeV, a three-dimensional sharp cutoff $\Lambda = 590$~MeV and
$G_0\Lambda^2 = 2.435$, corresponding to vacuum constituent masses
$M_{vac} = 400$~MeV. With these parameters we obtain reasonable values for
the pion mass, decay constant and the quark condensate in the vacuum:
$m_\pi = 140.2$~MeV, $f_\pi = 92.6$~MeV and 
$\ave{\bar u u} = (-241.5 \,\mathrm{MeV})^3$. 

\begin{figure}[t]
\begin{center}
\includegraphics*[bb = 108 380 485 519,height = 5.5cm]{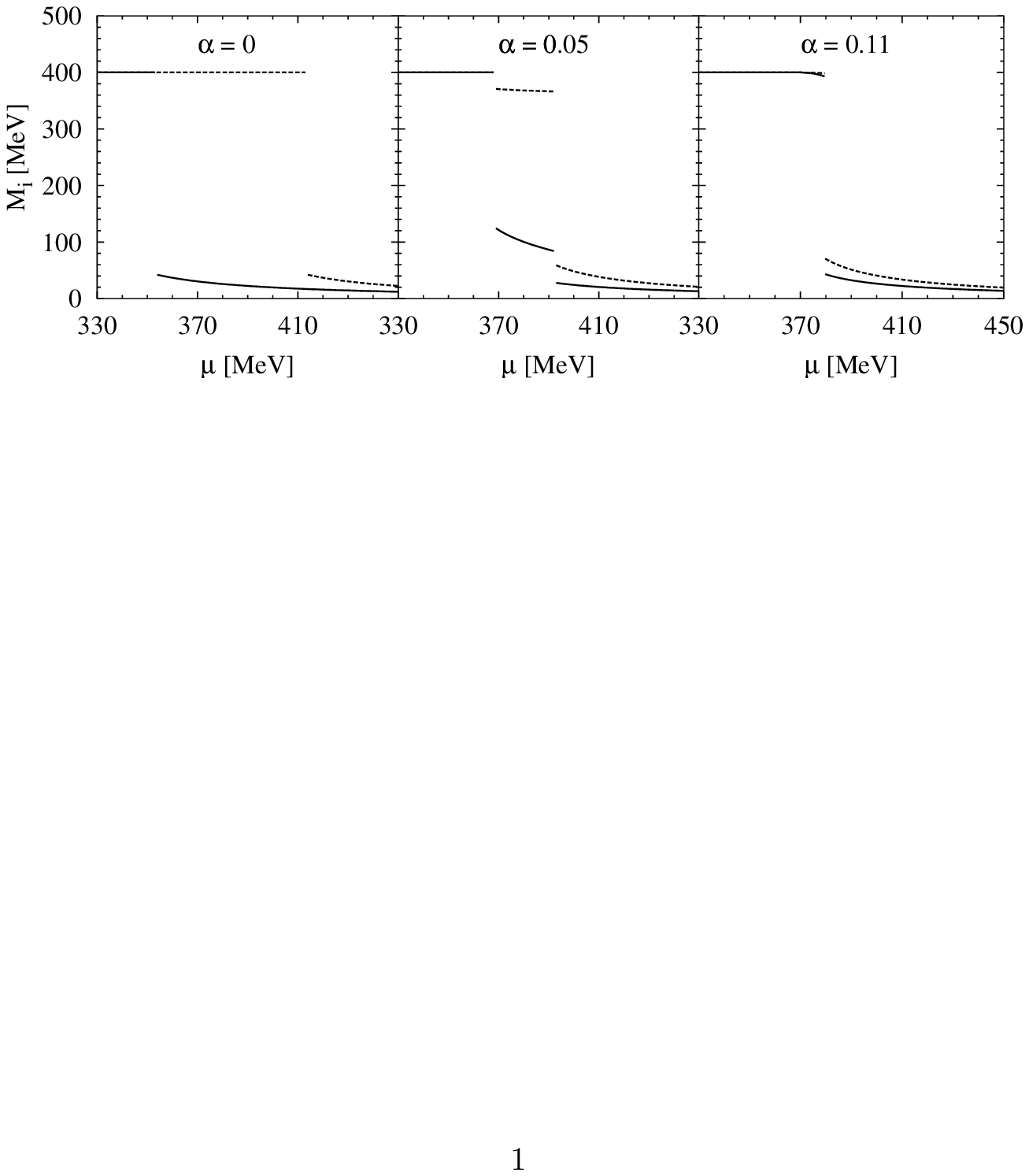}
\caption{Constituent quark masses $M_u$ (solid) and $M_d$ (dashed)
at $T=0$ 
as functions of quark number chemical potential $\mu$ for $\delta\mu = 30$ MeV 
and  $\alpha = 0$ (left), $\alpha = 0.05$ (center), and $\alpha = 0.11$ 
(right).
}
\label{masst0}
\end{center}
\end{figure}

We begin our discussion with the results at $T =
0$. In Fig.~\ref{masst0} we display the values of $M_u$ and $M_d$ as
functions of the quark number chemical potential $\mu$ for fixed  
$\delta\mu = 30$~MeV~\footnote{Following common practice 
(e.g., \cite{Bedaque02,KYT02,TK03})
we take a positive value of $\delta\mu$, although for the description of
heavy-ion collisions $\delta\mu<0$ would be more appropriate. However,
since changing the sign of $\delta\mu$ does only interchange the roles of up 
and down quarks, this does not alter our conclusions.}. 
The left panel corresponds to $\alpha = 0$, i.e., to
the case without flavor mixing. 
We observe two distinct phase transitions at $\mu = 353$~MeV for
the up quarks and at $\mu = 413$~MeV for the down quarks. This behavior
is easily understood when we recall that at $\alpha = 0$ the up and
down quark contributions to the thermodynamic potential completely decouple.
Hence, if we had plotted $M_u$ and $M_d$, in terms of the corresponding
{\it flavor} chemical potential $\mu_u$ and $\mu_d$, respectively,
we would have found two identical functions with a phase transition at
$\mu_f = 383$~MeV. This is basically the result reported in 
Refs.~\cite{KTV03,TK03}.  

Now we want to study the influence of a non-vanishing flavor mixing.
In the central panel of Fig.~\ref{masst0} we show the behavior of the 
constituent quark masses for $\alpha = 0.05$.
The situation remains qualitatively unchanged, i.e.,
we still find two distinct phase transitions.
However, because $M_d$ now also depends on $\phi_u$ (and thus on
$M_u$), and vice versa, both constituent masses drop at both critical
chemical potentials.  
Moreover, this small amount of flavor mixing already diminshes the
difference between the two critical quark number chemical potentials
considerably.
Finally, for $\alpha$ larger than a critical value  of $0.104$
we find only one {\it single} first-order phase transition.
This is illustrated in the right panel of  Fig.~\ref{masst0},
which corresponds to $\alpha = 0.11$.

\begin{figure}[t]
\begin{center}
\includegraphics*[bb = 108 380 485 519,height = 5.5cm]{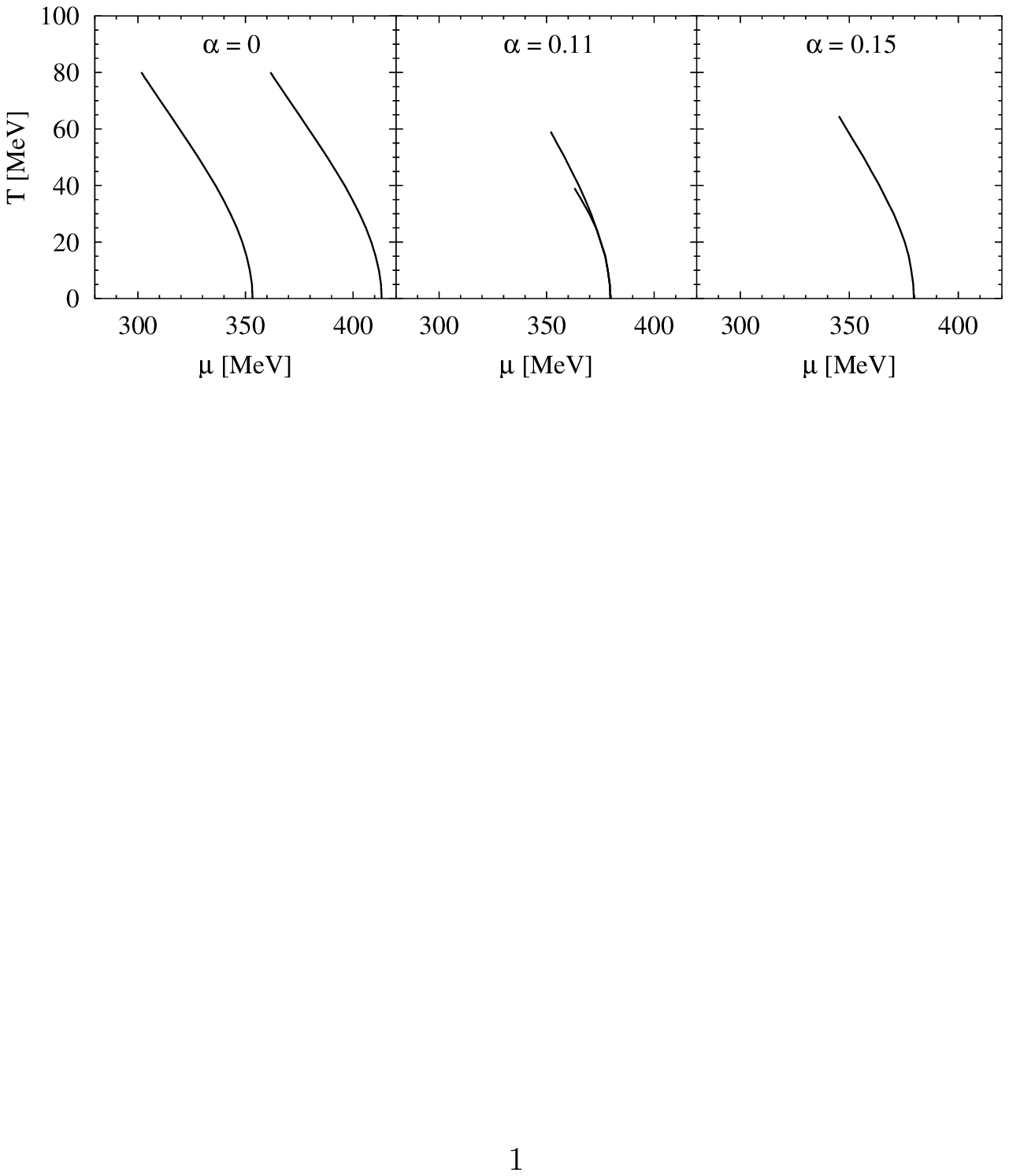}
\caption{Phase diagrams in the $\mu$-$T$-plane for 
$\delta\mu = 30$~MeV and $\alpha = 0$ (left), $\alpha = 0.11$ (center),
and $\alpha = 0.15$ (right). The lines correspond to first-order
phase boundaries which end in second-order endpoints.}
\label{phase}
\end{center}
\end{figure}

Next, we extend our analysis to non-vanishing temperature.
The phase diagrams in the $\mu$-$T$ plane for fixed $\delta\mu = 30$~MeV
and three different values of $\alpha$ are shown in Fig.~\ref{phase}.
At $\alpha = 0$ (left panel) we qualitatively reproduce the results 
discussed in Refs.~\cite{KTV03,TK03}, i.e., two separate first-order phase 
boundaries which end in two second-order endpoints.
Again, since for $\alpha = 0$ the up and down quarks
decouple completely, we would obtain two identical phase diagrams
if we plotted the phase structure of flavor $f$ in the $\mu_f$-$T$ plane.
In the central panel of Fig.~\ref{phase} we consider $\alpha = 0.11$,
i.e., slightly larger than the critical value $\alpha^c(T=0) = 0.104$
for a single phase transition at $T=0$.
Accordingly, there is only one phase boundary at low temperatures,
but at $T = 25$~MeV it splits into two lines which end at two 
different second-order endpoints. 
The two branches are, however, very close to each other and already
at $\alpha = 0.12$ we find only one phase boundary with a single endpoint.
This is illustrated by the diagram on the right, which corresponds
to $\alpha = 0.15$.

In our example a rather small amount of flavor mixing is sufficient
to remove the existence of the second phase transition:
Of course, there {\it must} be a single phase transition at $\alpha = 0.5$,
where $M_u$ and $M_d$ are equal (see \eq{masses}). 
(This was the case studied in Ref.~\cite{KYT02}.) However, the critical
value $\alpha^c \simeq 0.12$ we found in our example is much smaller.
At $T=0$, a rough, but perhaps more general estimate for the critical
$\alpha$ can be obtained from the observation that the phase transition 
takes place when 
the chemical potential of quark $f$ comes close to its constituent
mass, i.e., $\mu_f \approx M_i$. Applying this condition to the 
$u$ quark we expect the first phase transition to take place
at $\mu_u \approx M_{vac}$, i.e, at $\mu \approx M_{vac} - \delta\mu$.
At this point $M_u$ drops and, according to \eq{masses}
$M_d$ drops as well. Neglecting the current quark mass, we find  
\beq
    M_d \;\approx\; -(1-\alpha)\,4G_0\, \phi_d 
        \;\lsim\; (1-\alpha)\, M_\mathit{vac}~.
\label{md}
\eeq
If this value becomes smaller than the
value of $\mu_d$, we expect also the down quarks to exhibit a phase
transition. Hence, we estimate 
\beq 
\alpha^c(T=0) \lsim \frac{ 2\delta\mu}{M_\mathit{vac}}~.  
\label{ac}
\eeq
Note that this estimate would not be affected by a possible restoration
of the $U_A(1)$ symmetry at the phase boundary. 
Obviously, if $G_2$ goes to zero, $M_d$ would drop as well.

For our example, \eq{ac} gives $\alpha^c(T=0) \lsim 0.15$.
Comparing this value with the numerical result $\alpha^c(T=0) = 0.104$, we
see that \eq{ac} is a quite conservative estimate. This is easily
understood, since in the second step of \eq{md} we have neglected
the fact, that $\phi_d$ also becomes smaller.  
Our estimate does also not include the observation, that
the critical chemical potential for the first phase transition {\it rises}
with $\alpha$.
In any case, we have to admit that our arguments cannot explain 
quantitatively why \eq{ac} seems to hold even for temperatures approaching
the critical endpoint where the quark masses do no longer drop discontinuously.

At this point one can ask, which value of $\alpha$ is ``realistic''.
To answer this question it is helpful to have a look at the 
3-flavor NJL model, where the strength of the 't Hooft interaction
has been fitted by several authors to describe the 
$\eta$-$\eta'$-splitting~\cite{Rehberg,Kunihiro}.
For three flavors, the 't Hooft determinant is a six-point 
interaction~\cite{tHooft}, and the constituent quark masses in a 
3-flavor NJL model are given by
\beq
     M_i = m_i - 4\,G\,\phi_i +2 \,K\,\phi_j\phi_k~,
     \qquad i\neq j \neq k \neq i \in \{u,d,s\}~.
\label{masses3}
\eeq    
When we compare this with \eq{masses} we can identify $G_1 = G$
and $G_2 = -\frac{1}{2} K \phi_s$ and thus
\beq
    \alpha = \frac{-\frac{1}{2} K \phi_s}{G -\frac{1}{2} K \phi_s}~.
\eeq
If we take, for instance, the values of Ref.~\cite{Rehberg}, 
$\Lambda = 602.3$~MeV, $G\Lambda^2 = 1.835$, $K\Lambda^5 = 12.36$, and 
$\phi_s = (-257.7\, \mathrm{MeV})^3$, we find $\alpha \simeq 0.21$.
For the parameters of Ref.~\cite{Kunihiro} we get a somewhat smaller
value, $\alpha \simeq 0.16$. 
On the other hand, the success of the instanton liquid model to describe
hadronic correlation functions~\cite{SS98} would suggest that ${\cal L}_2$ is 
the dominant part of our Lagrangian, i.e., $\alpha\simeq 1$.
Anyway, in all cases we would find
only one phase transition for $\delta\mu = 30$~MeV. 
Typical values of $|\delta\mu|$ in heavy-ion collisions are likely to be
smaller than that. 
(A simple estimate, assuming the density ratio  
$n_u : n_d = 290 : 334$ as in $^{208}$Pb, and the approximate relation
$n_u : n_d \approx (\mu_u : \mu_d)^3$ yields $\delta\mu \approx -10$~MeV
for $\mu = 400$~MeV.
Empirically, one finds $\mu_I = 2\delta\mu = -5$~MeV at chemical freeze-out
for Pb-Pb collisions at SPS~\cite{BMHS99} and $\mu_I = -12$~MeV
for Si+Au collisions at AGS~\cite{heppediplom}.)

However, before drawing quantitative conclusions,
we should be aware of several shortcomings of the present model.
First, the description of the ``hadronic phase'' as a gas of
quarks, rather than hadrons, is certainly unrealistic. 
In fact, it is quite obvious that any prediction of the critical endpoint(s)
in non-confining mean-field models should not be trusted.
(Lattice calculations, although not yet generally accepted, indicate
that the critical endpoint of QCD at $\delta\mu = 0$
is located at $\mu_B = 3\mu \simeq 725 \pm 35$~MeV and 
$T = 160 \pm 3.5$~MeV~\cite{FoKa02},
i.e., at lower chemical potential and higher temperature than our or
other NJL results, e.g.,~\cite{TK03,AsYa89}.) 
Moreover, as already mentioned, we have neglected the possibilities
of pion and diquark condensation. 
Since pion condensation, although interesting by itself, does only 
occur for $\delta\mu \gtrsim 70$~MeV, it is irrelevant for the
present discussion. A possible diquark condensation in the
2SC phase~\cite{RaWi00} would further favor a single phase transition, 
since this phase requires an approximately equal number of up and down
quarks. An extension of the present model to include 
these possibilities is straight forward.

Keeping the limitations of our model in mind, our results
show that flavor-mixing effects cannot be
neglected in the discussion of the phase diagram. The pure existence
of these effects is related to instantons and the $U_A(1)$-anomaly of
QCD. Of course their magnitude is a matter of debate,
but they probably cancel the interesting phenomena discussed in
Refs.~\cite{KTV03,TK03}. 
\\[2mm]
{\bf Acknowledgments:}\\
We thank P. Braun-Munzinger for pointing out Ref.~\cite{heppediplom}
to us. 
M.O. acknowledges support from the Alexander von
Humboldt foundation as a Feodor-Lynen fellow.


\begin{thebibliography}{99}
\bibitem{RaWi00} K. Rajagopal and F. Wilczek, hep-ph/0011333,
                 and references therein.
\bibitem{Alford01} M. Alford, Ann. Rev. Nucl. Part. Sci. 51 (2001) 131.
\bibitem{K01} F. Karsch, AIP Conf. Proc. 602 (2001) 323; \\
J.B. Kogut, hep-lat/0208077.
\bibitem{HJSSV98} M.A. Halasz, A.D. Jackson, R.E. Shrock,
M.A. Stephanov, and J.J. Verbaarschot, Phys. Rev. D 58 (1998) 096007.
\bibitem{SRS99} M. Stephanov, K. Rajagopal, and E. Shuryak,
Phys. Rev. D 60 (1999) 114028.
\bibitem{Bedaque02} P.F. Bedaque, Nucl. Phys. A 697 (2002) 569.
\bibitem{ABR01} M.G. Alford, J.A. Bowers, and K. Rajagopal,
                Phys. Rev. D 63 (2001) 074016.
\bibitem{KYT02} O. Kiriyama, S. Yasui, and H. Toki,
                Int. J. Mod. Phys. E 10 (2002) 501.
\bibitem{SoSt01} D.T. Son and M. Stephanov, Phys. Rev. Lett. 86 (2001) 592.
\bibitem{KoSi02} J.B. Kogut and D.K. Sinclair, Phys. Rev. D 66 (2002), 014508;
                 ibid. 034505.
\bibitem{KTV03} B. Klein, D. Toublan, and J.J.M. Verbaarschot, hep-ph/0301143.
\bibitem{TK03} D. Toublan, J.B. Kogut, hep-ph/0301183.
\bibitem{AsYa89} M. Asakawa and K. Yazaki, Nucl. Phys. A 504 (1989) 668.
\bibitem{Klevansky}  S.P. Klevansky, Rev. Mod. Phys. 64 (1992) 3.
\bibitem{tHooft} G. 't Hooft, Phys. Rev. D 14 (1976) 3432;
                 Phys. Rep. 142 (1986) 357.
\bibitem{NJL} Y. Nambu and G. Jona-Lasinio, Phys. Rev. 122 (1961) 345;
              124 (1961) 246.
\bibitem{Rehberg} P. Rehberg, S.P. Klevansky and J. H\"ufner,
                  Phys. Rev. C 53 (1996) 410.
\bibitem{Kunihiro} T. Kunihiro, Phys. Lett. B 219 (1989) 363.
\bibitem{SS98} T. Sch\"afer and E. Shuryak, Rev. Mod. Phys. 70 (1998) 323.
\bibitem{BMHS99} P. Braun-Munzinger, I. Heppe, and J. Stachel,
                 Phys. Lett. B 465 (1999) 15.
\bibitem{heppediplom} I. Heppe, Diplomarbeit, Universit\"at Heidelberg
1998, unpublished\\ 
(http://www.physi.uni-heidelberg.de/physi/ceres/publications/IHeppe-diplom.ps).
\bibitem{FoKa02} Z. Fodor and S.D. Katz, JHEP 0203 (2002) 014.
\end{thebibliography}
\end{document}